\documentclass[
 reprint,
 superscriptaddress,
 groupedaddress,
 showpacs,preprintnumbers,
 nobibnotes,
 amsmath,amssymb,
 aps,
 floatfix,
]{revtex4-1}

\usepackage{hyperref}

\usepackage{graphicx}
\usepackage{color}
\usepackage{dcolumn}
\newcolumntype{d}[1]{D{.}{.}{#1}}

\newcommand{\paralel}{\mathbin{\!/\mkern-5mu/\!}}
\newcommand{\psiE}{\psi_{\scalebox{.95}{$\scriptscriptstyle \mathbf{E}$}}}
\newcommand{\rhoE}{\rho_{\scalebox{.95}{$\scriptscriptstyle \mathbf{E}$}}}

\begin{document}

\title{
Electrical tuning of spin splitting in Bi-doped ZnO nanowires
}

\author{Mehmet Aras}
\author{\c{C}etin K{\i}l{\i}\c{c}}\email{cetin\_kilic@gtu.edu.tr}
\affiliation{Department of Physics, Gebze Technical University, 41400 Gebze Kocaeli, Turkey}


\begin{abstract}
\centerline{\sl Published version available at \url{https://doi.org/10.1103/PhysRevB.97.035405}}
\vspace*{6pt}
The effect of applying an external electric field
  on 
  doping-induced spin-orbit splitting of the lowest conduction-band states in a bismuth-doped zinc oxide nanowire
  is studied 
  by performing 
  electronic structure calculations
  within the framework of density functional theory.
It is demonstrated that
  spin splitting in Bi-doped ZnO nanowires could be tuned and enhanced electrically
  via control of the strength and direction of the applied electric field,
  thanks to the non-uniform and anisotropic response of the ZnO:Bi nanowire to external electric fields.
The results reported here indicate that
  a single ZnO nanowire doped with a low concentration of Bi
  could function as a spintronic device,
  operation of which is controlled by applied lateral electric fields.
\end{abstract}

\maketitle

In the presence of noncentrosymmetric electric fields,
  the spin-orbit (SO) interaction leads to a $k$-dependent splitting of electronic states,
  enabling electrical control 
  of the spin-split states in spintronic devices \cite{rashba03,sahoo05,nowack07}.
The development of a class of spintronic materials
  is thus facilitated by
  engineering (or exploiting) inversion asymmetries to generate \textit{intrinsic} electric fields \cite{ishizaka11,disante13,liu13,kilic15,cheng16}
  as well as devising architectures in which \textit{external} electric fields are used \cite{nitta97,liang12,gong13}.
It has recently been proposed that
  surface deposition \cite{calleja15} and doping \cite{aras17} with \textit{heavy} elements
  could also be used to develop materials with spintronic functionalities.
In particular, the predictions of Ref.~\onlinecite{aras17} 
  show that doping a \textit{light} semiconducting (ZnO) nanowire with 
  a heavy element (Bi)
  leads to linear-in-$k$ splitting of the conduction-band (CB) states through SO interaction.
It is thus anticipated that
  a \textit{single} ZnO nanowire doped with a low concentration of Bi could function as a spintronic device.
The objective of the present paper is to investigate
  if the spintronic properties of a ZnO:Bi nanowire
  could be tuned or \textit{enhanced} electrically.
In our previous paper\cite{aras17}, we showed that
  the $k$-dependent SO splitting in ZnO:Bi nanowires could be tuned by adjusting the dopant concentration.
Here we demonstrate that
  applying external electric fields provides an effective means to enhance
  the linear-in-$k$ SO splitting of the CB states in ZnO:Bi nanowires.
We find that
  the SO splitting energy could be made to have a \textit{superlinear} increase
  with increasing electric field strength $E$,
  which is mediated by controlling the direction of the applied electric field $\mathbf{E}$.
This is found to be facilitated by
 the \textit{non-uniform} and \textit{anisotropic} response of the ZnO:Bi nanowire
 to external electric fields.
The latter remind the (converse) piezoelectric response of
  undoped ZnO nanowires \cite{he07, agrawal11, broitman13} and microbelts \cite{hu09}.
On the other hand, our results also indicate
  that the presence of the substitutional Bi dopant on the ZnO nanowire surface reduces the amount of deformation of the nanowire under an electric field.

As long as single Co-doped ZnO nanowires and nanorods
  have been produced and characterized \cite{liang09,segura11,ko12},
  we think that
  the realization of a single ZnO:Bi nanowire
  is \textit{not} beyond the reach of current capabilities,
  although differences between cobalt and bismuth (e.g., bismuth's larger ionic radius \cite{shannon69} and lower solubility \cite{smith89} in bulk ZnO)
  should be taken into consideration.
Since experimental studies on single Bi-doped ZnO nanowires were not available (to our knowledge),
  the realization and stability of a single ZnO:Bi nanowire were examined theoretically in our previous publications \cite{kilic16,aras17}
  with the aid of \textit{defect} calculations and finite-temperature \textit{ab initio} molecular dynamics simulations,
  as will be discussed in the Appendix.

The findings reported here were obtained
  via electronic structure calculations
  performed within framework of the density functional theory (DFT)
  by employing periodic supercells.
Although the supercells were in practice subject to the Bloch periodicity condition in all directions,
  the supercell dimensions perpendicular to the nanowire axis were set to be significantly larger than the nanowire diameter
  in order to create a vacuum region (of thickness larger than 15~\textup{\AA}) that avoid interactions between the nanowire and its periodic images.
We used the Vienna \textit{ab initio} simulation package \cite{kresse96} (VASP)
  together with its projected-augmented-wave potential database \cite{kresse99},
  adopting the rotationally-invariant DFT+$U$ approach \cite{dudarev98}
  in combination with the Perdew-Burke-Ernzerhof exchange-correlation functional \cite{perdew96},
  and taking the SO coupling into account as implemented \cite{hobbs00,steiner16} in the VASP code.
The 2$s$ and 2$p$,
               3$d$ and 4$s$, and
               6$s$ and 6$p$
  states were treated as valence states for oxygen,
                               zinc, and
                               bismuth, respectively.
Plane wave basis sets with a kinetic energy cutoff of 400 eV
  were used to represent the electronic states.
In test calculations \cite{aras17}
  the kinetic energy cutoff was increased by 10~\% and the change in the SO splitting energies
  turned out to be  smaller than 0.5~\%.
The value of Hubbard $U$ was set to 7.7 eV \cite{kilic16},
  which was applied to the Zn 3$d$ states.
The DFT+$U$ approach was preferred over the standard (semilocal) DFT calculations
  in order to reduce the underestimation of the $d$ state binding energies \cite{aras14}.
We applied \textit{lateral} electric fields in the $x$ or $y$ directions
  of varying strength $E$ (from 0.1 to 0.5 eV/\textup{\AA}\ with an increment of 0.1 eV/\textup{\AA}),
  orienting the nanowire axis along the $z$ direction.
It should be reminded that
  VASP handles the external electric fields
  by introducing artificial dipole sheets in the middle of the vacuum regions
  in the supercell, cf. Ref.~\onlinecite{feibelman01}.
Structural optimizations were
  performed for each atomic configuration,
  separately for each given $\mathbf{E}$,
  by minimizing the total energy
  until the maximum value of residual forces on atoms was reduced to be smaller than $10^{-2}$~eV/\textup{\AA},
  using the $\Gamma$-point for sampling the supercell Brillouin zone (BZ).
We determined \cite{aras17}
  an error bar of 0.2 meV for the energy per ZnO unit
  owing to the BZ sampling achieved through zone folding.
Convergence criterion for the electronic self-consistency was set up to
  10$^{-6}$~eV and 10$^{-8}$~eV in structural optimizations and electronic structure calculations, respectively.

\begin{figure}
\centering
  \includegraphics[width=0.500\textwidth]{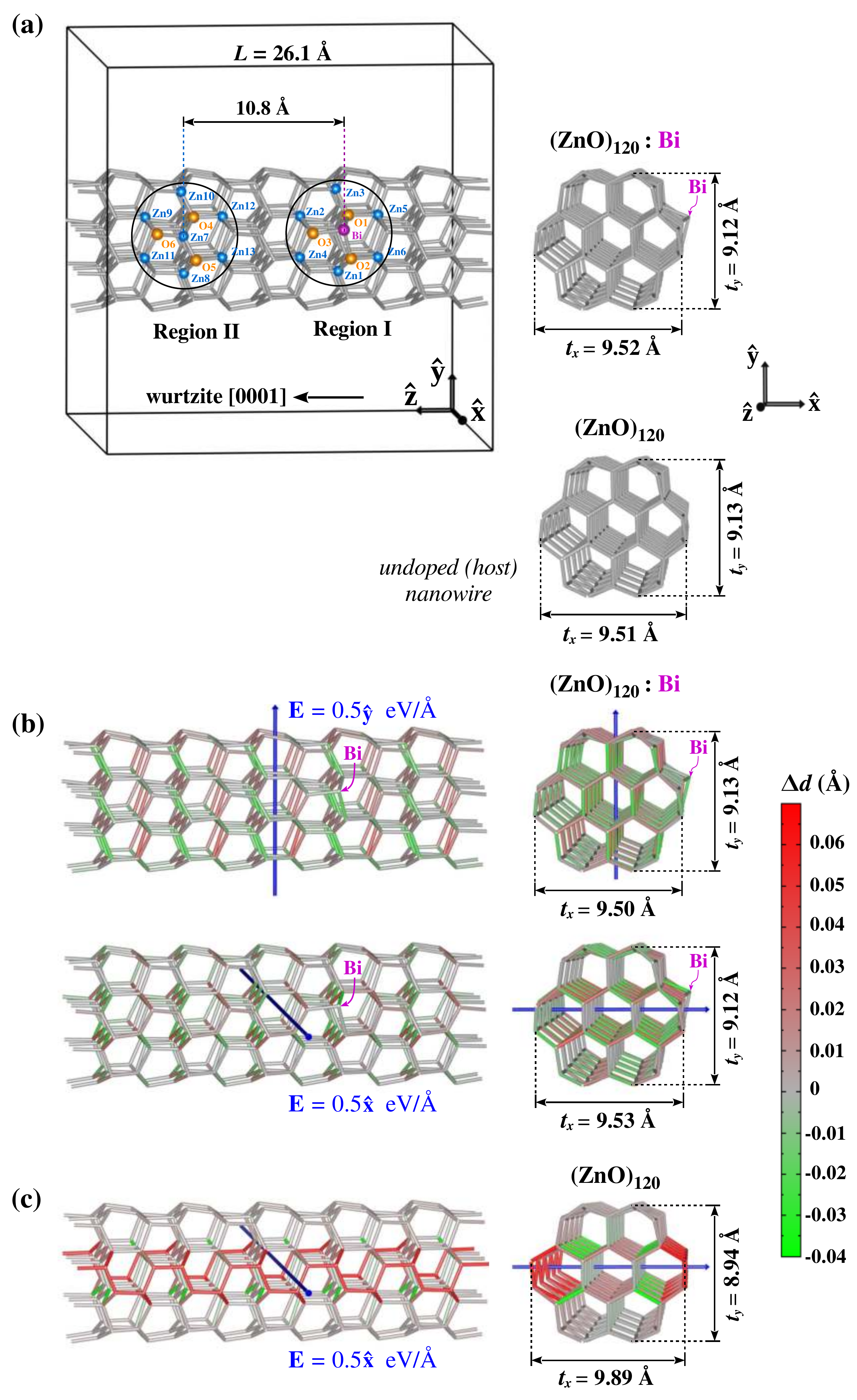}
  \caption{The equilibrium atomic configuration for
           the Bi-doped nanowire (a) in the absence of an external electric field
            and (b) for $\mathbf{E}=0.5\hat{\mathbf{y}}$ and $\mathbf{E}=0.5\hat{\mathbf{x}}$~eV/\textup{\AA}.
           (c) The same for the undoped (i.e., host) nanowire for $\mathbf{E}=0.5\hat{\mathbf{x}}$~eV/\textup{\AA}.
           The host nanowire for $\mathbf{E}=\mathbf{0}$ is also included in (a).
           The sticks in (b and c) representing the Zn-O bonds are colored to reflect the electric-field-induced changes $\Delta d$ in the bond lengths.
           The thicknesses $t_x$ and $t_y$ are indicated.
          }
  \label{f:as}
\end{figure}

\begin{figure*}
\centering
  \includegraphics[width=1.000\textwidth]{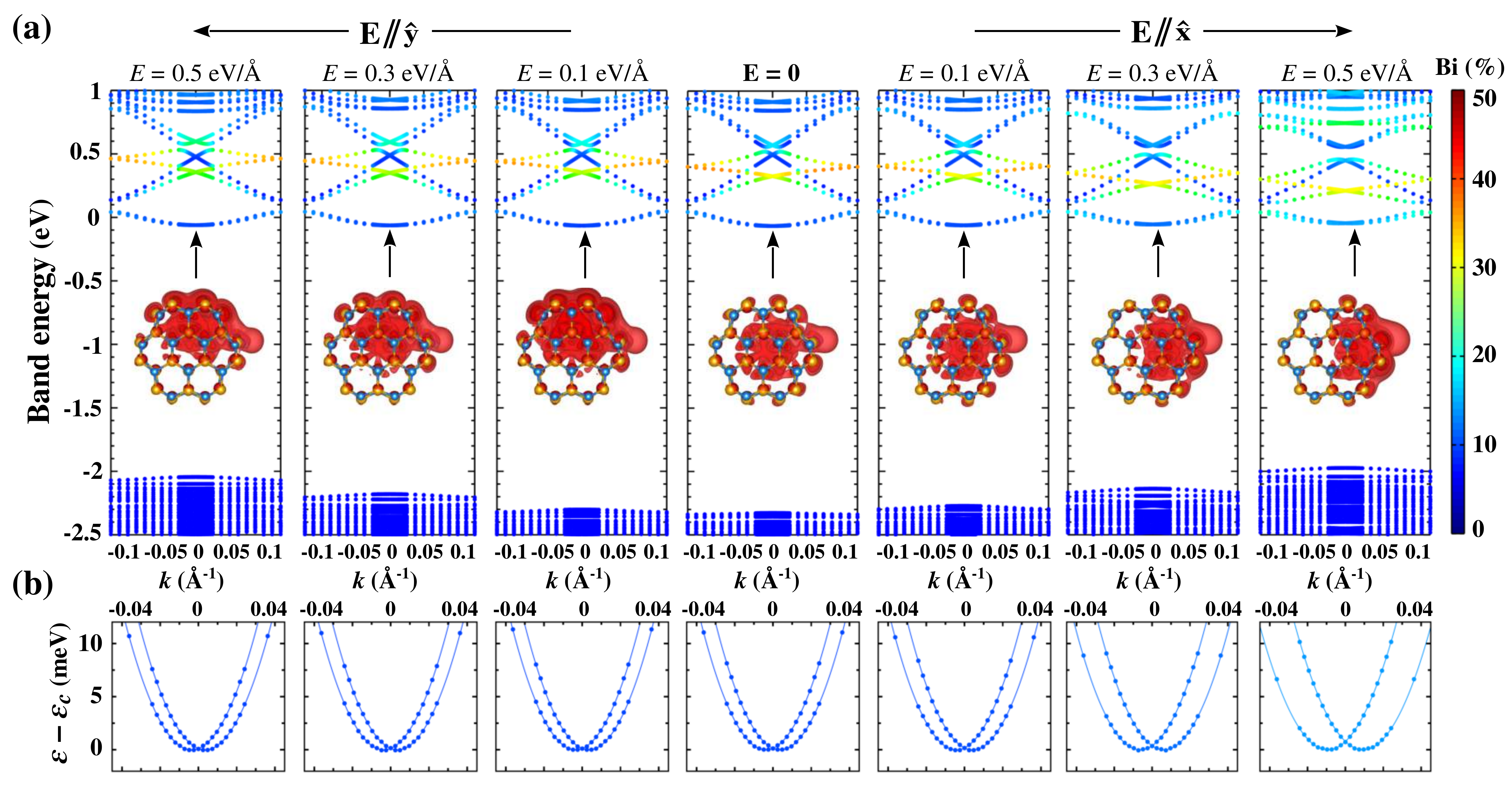}
  \caption{(a) The electronic energy bands of the (ZnO)$_{120}$:Bi nanowire under applied electric fields,
           which are colored to reflect the percent contribution from Bi to the electronic states.
           The Fermi level is set as the zero of energy.
           The vertical arrows point to the CBM.
           The red isosurfaces in the insets of the upper panels represent the CBM state charge densities for the isovalue of $1 \times 10^{-4}$~\textup{\AA}$^{-3}$.
           (b) Close-up views showing the spin-orbit splitting of the \textit{two} lowest conduction bands.
           }
  \label{f:bs}
\end{figure*}

Figure~\ref{f:as}(a) shows the equilibrium atomic configuration
  for a Bi-doped ZnO nanowire in the absence of an external electric field,
  where the Bi dopant substitutes Zn at a surface site
  of the host ZnO nanowire.
We refer to the Appendix
  for a discussion of issues concerning the realization of this configuration.
The equilibrium atomic configuration
  for the undoped (host) nanowire is also shown in Fig.~\ref{f:as}(a),
  which was used in former theoretical studies by others (e.g., Refs.~\onlinecite{fan07,agrawal11}) as well as the authors \cite{kilic16,aras17}.
The wire thicknesses $t_x$ and $t_y$ are indicated in Fig.~\ref{f:as}(a)
  for both the doped and undoped nanowires.
Comparing these thicknesses,
  it is clear that
  the incorporation of Bi causes insignificant deformation in the wire morphology,
  which means that the accommodation of Bi induces mostly local relaxations.
It should also be pointed out that
  the host nanowire's thickness is smaller than the experimental diameter ($D$) values measured for
  high-aspect-ratio ZnO nanorods with $D \ge 2.2$~nm \cite{yin04}
  and thin ZnO nanowires with $D \ge 4.1$~nm \cite{stichtenoth07}.
The variation of the results with respect to the nanowire's thickness
  was studied in Ref.~\onlinecite{aras17} (in the absence of an external electric field),
  which will not be done here.

The equilibrium configurations in the presence of electric fields are
  given in Figs.~\ref{f:as}(b) and S1 (see Ref.~\onlinecite{supmat}).
Since the electric-field-induced changes in the atomic positions
  are not visible in the scale of the figures,
  the sticks representing the Zn-O bonds are colored
  to reflect the electric-field-induced changes $\Delta d$ in the bond lengths.
Hence, 
  the structural response of the ZnO:Bi nanowire to the applied electric fields
  could be inferred from
  the distributions of red and green sticks 
  (representing the elongated and shrunk bonds, respectively).
It is clear in Figs.~\ref{f:as}(b) and S1 (see Ref.~\onlinecite{supmat})
  that the electric-field-induced structural changes occur all around the nanowire.
Nevertheless,
  a comparison between the bond lengths in Regions I and II depicted in Fig.~\ref{f:as}(a),
  the values of which are provided in Table~S1 (see Ref.~\onlinecite{supmat}),
  reveals that
  the structural changes are more pronounced in the vicinity of the dopant.
For example, the O1$-$Bi (O2$-$Bi) bond located in Region I exhibits the greatest shrinkage
  for $\mathbf{E} \paralel \hat{\mathbf{x}}$ ($\mathbf{E} \paralel \hat{\mathbf{y}}$),
  the degree of which is proportional to $E$.
The respective bonds in Region II (far from the dopant),
  i.e., the O4$-$Zn7 and O5$-$Zn7 bonds, however
  exhibit considerably smaller shrinkage.

As mentioned above,
  the presence of the Bi dopant on the ZnO nanowire surface reduces 
  the amount of deformation in the wire morphology under a lateral electric field.
This could be seen by comparing the structures in Figs.~\ref{f:as}(b) and (c).
For the doped nanowire,
  the $t_x$ and $t_y$ values in Fig.~\ref{f:as}(b) are \textit{little} different
  from those in Fig.~\ref{f:as}(a).
In contrast,
  there is a \textit{significant} increase (decrease) in
  the $t_x$ ($t_y$) of the undoped nanowire
  as a result of applying $\mathbf{E} = 0.5 \hat{\mathbf{x}}$~eV/\textup{\AA},
  yielding a noticeable modification in the wire cross-section
  since the ratio $t_x/t_y$ changes from 1.04 to 1.11.
The distribution of the stretched (shrunk) bonds, cf. red (green) sticks, in Fig.~\ref{f:as}(c)
  is also noticeably different than that in Fig.~\ref{f:as}(b),
  which reveals the microscopic origin of the increase (decrease) in $t_x$ ($t_y$).
It should be noticed that
  applying a lateral electric field make the Zn-O bonds that are aligned with the wire axis elongate
  according to our prediction,
  which is in accordance with the response of a ZnO microbelt \cite{hu09} to an applied electric field perpendicular to its $c$-axis.

The electronic energy bands of the (ZnO)$_{120}$:Bi nanowire,
  calculated
  for $\mathbf{E}=$
  0.5$\hat{\mathbf{y}}$, 0.4$\hat{\mathbf{y}}$, 0.3$\hat{\mathbf{y}}$, 0.2$\hat{\mathbf{y}}$, 0.1$\hat{\mathbf{y}}$,
  $\mathbf{0}$,
  0.1$\hat{\mathbf{x}}$, 0.2$\hat{\mathbf{x}}$, 0.3$\hat{\mathbf{x}}$, 0.4$\hat{\mathbf{x}}$, and 0.5$\hat{\mathbf{x}}$~eV/\textup{\AA},
  are shown in Figs.~\ref{f:bs}(a) and S2(a) (see Ref.~\onlinecite{supmat})
  where the symbols are colored to reflect the percent contribution from Bi to the electronic states.
The coloring is accomplished
  by computing the contributions from the Zn, O, and Bi atoms
  that are obtained by projecting the state wave functions onto spherical harmonics within a sphere around each atom.
The vertical arrows point to the conduction-band minimum (CBM) that occurs at $k=k_c$.
It is seen that Bi-derived states occur as resonances in the conduction band,
  energies of which get lowered (remain roughly constant)
  for $\mathbf{E} \paralel \hat{\mathbf{x}}$ ($\mathbf{E} \paralel \hat{\mathbf{y}}$).
The CBM state charge density $\rhoE(\mathbf{r})=|\psiE(\mathbf{r})|^2$
  is noticeably distorted
  \textit{in a directed manner} as imposed by the direction of the applied electric field $\mathbf{E}$,
  which is inferred from the isosurfaces given as insets in Fig.~\ref{f:bs}(a).
Whereas Bi contribution to the CBM wave function $\psiE$ decreases slowly with $E$ for $\mathbf{E} \paralel \hat{\mathbf{y}}$,
  applying $\mathbf{E}$ in the $x$ direction makes the CBM wave function
  have a higher contribution from Bi, in proportionality with $E$.
Thus an important effect of applying an external electric field is to vary the Bi contribution to the lower CB states.

\begin{figure}
\centering
  \includegraphics[width=0.500\textwidth]{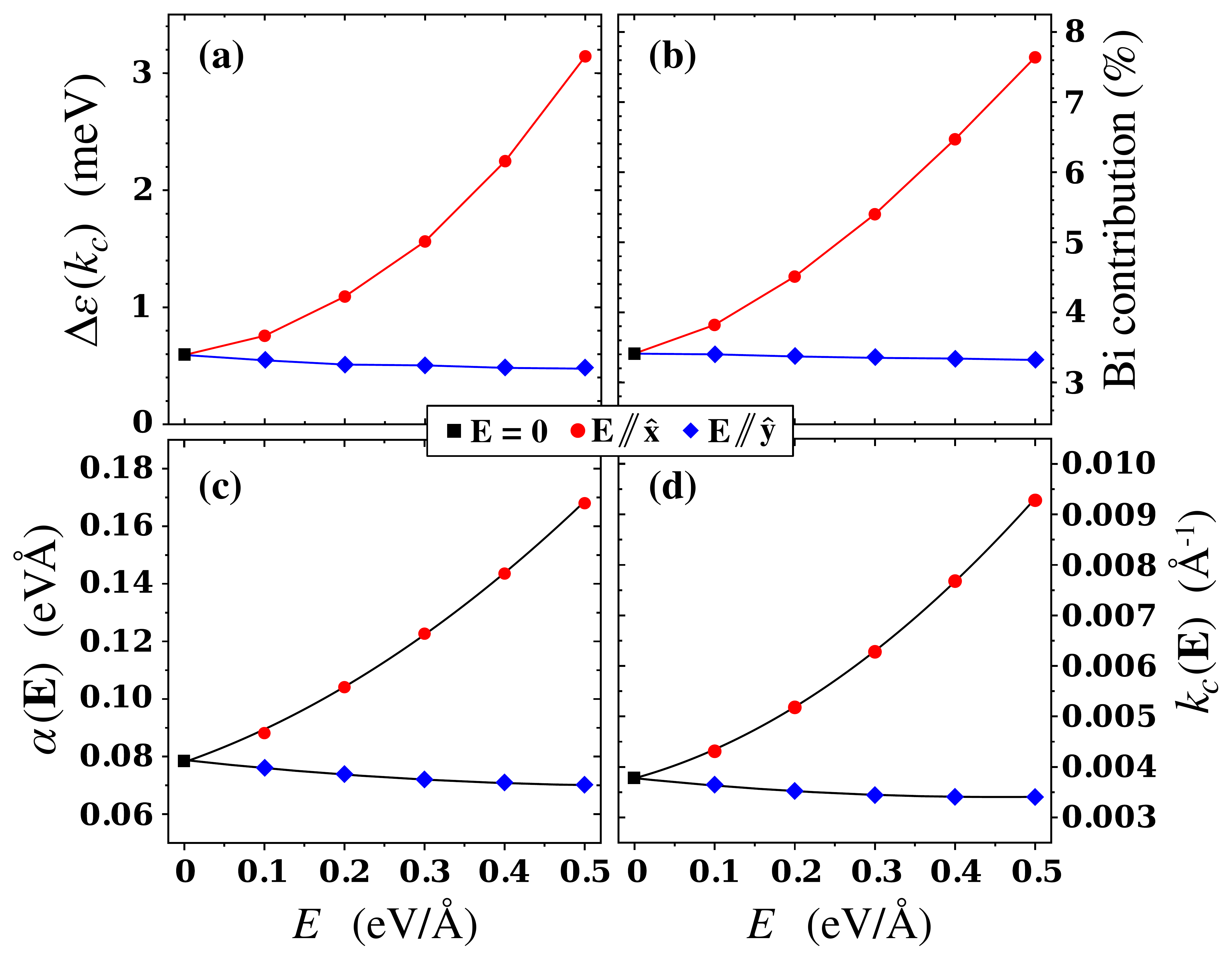}
  \caption{The variation of the (a) splitting energy $\Delta \varepsilon(k_c)$,
                                (b) Bi contribution to the CBM wave function,
                                (c) linear coefficient $\alpha$, and
                                (d) momentum offset $k_c$
          with the external electric field strength $E$.}
  \label{f:var}
\end{figure}

Figures~\ref{f:bs}(b) and S2(b) (see Ref.~\onlinecite{supmat}) display close-up views of the \textit{two} lowest (spin-split) conduction bands,
  where the bands are shifted by subtracting the lowest eigenvalue $\varepsilon_c$ of the conduction band from the band energies $\varepsilon$.
These bands are partially occupied
  since the aforementioned Bi-derived resonant states in the conduction band are empty,
  reflecting the donor behavior of Bi in ZnO nanowires \cite{kilic16}.
The dispersion of the spin-split bands in Fig.~\ref{f:bs}(b)
  is \textit{accurately} described by
\begin{equation}
  \varepsilon_{\pm}(k)=\frac{\hbar^2}{2m^{\ast}}k^2 \pm \alpha k,
  \label{e:disp}
\end{equation}
represented by the solid curves in each panel.
Although Eq.~(\ref{e:disp}) is of the same form of the Bychkov-Rashba expression \cite{bychkov84},
  the electrons filling the $\varepsilon_{\pm}$ bands do \textit{not} form a two-dimensional electron gas,
  as seen from the insets of Figs.~\ref{f:bs}(a) and S2(a) (see Ref.~\onlinecite{supmat}).
The latter applies to the case of \textit{zero} electric field,
  as discussed in detail in Ref.~\onlinecite{aras17}.
It is clear that
  the splitting energy $\Delta \varepsilon(k) =\varepsilon_{+}(k)-\varepsilon_{-}(k)$ increases (decreases slowly) 
  as $E$ increases for $\mathbf{E} \paralel \hat{\mathbf{x}}$ ($\mathbf{E} \paralel \hat{\mathbf{y}}$).
This is the same trend for the Bi contribution to the CBM wave function, 
  as noted in the preceding paragraph.
Accordingly
  the variations of $\Delta \varepsilon(k_c)$
  and the Bi contribution to $\psiE$
  with $E$ follow the same trend,
  as seen in Figs.~\ref{f:var}(a) and \ref{f:var}(b).
This means that increasing the Bi contribution to the lower CB states
  leads to the enhancement of spin-orbit splitting of those states.
It should also be pointed out that $\mathbf{E}$ induces
  a \textit{much smaller} splitting in the \textit{undoped} nanowire
  owing to the absence of the Bi contribution.
This is illustrated in Fig.~S3 (see Ref.~\onlinecite{supmat})
  where the linear coefficient $\alpha = 0.02$~eV\textup{\AA} in the absence of the heavy element Bi,
  which should be compared to the respective value of $\alpha = 0.17$~eV\textup{\AA} in the presence of the Bi dopant.

The splitting energy plotted in Fig.~\ref{f:var}(a)
  is given by $\Delta \varepsilon(k_c) =2\alpha k_c$
  according to Eq.(\ref{e:disp}).
Thus the variation of the linear coefficient $\alpha$ and the momentum offset $k_c$
  with $\mathbf{E}$ is studied in Figs.~\ref{f:var}(c) and \ref{f:var}(d), respectively,
  where the black curves represent the parameterization according to
\begin{eqnarray}
  \alpha(\mathbf{E})&=&
  \begin{cases}
    \alpha_{\text{\tiny 0}} + 0.097 E + 0.168 E^2 & \text{if}\ \mathbf{E} \paralel \hat{\mathbf{x}} \\
    \alpha_{\text{\tiny 0}} - 0.030 E + 0.026 E^2 & \text{if}\ \mathbf{E} \paralel \hat{\mathbf{y}},
  \end{cases}
  \\
  k_c(\mathbf{E})&=&
  \begin{cases}
    k_{\text{\tiny 0}} + 0.0044 E + 0.0134 E^2 & \text{if}\ \mathbf{E} \paralel \hat{\mathbf{x}} \\
    k_{\text{\tiny 0}} - 0.0016 E + 0.0179 E^2 & \text{if}\ \mathbf{E} \paralel \hat{\mathbf{y}}.
  \end{cases}
\end{eqnarray}
Here the units of $\alpha$, $k_c$ and $E$ are eV\textup{\AA}, \textup{\AA}$^{-1}$ and eV/\textup{\AA}, respectively;
  $\alpha_{\text{\tiny 0}}=\alpha(\mathbf{0})=0.079$~eV\textup{\AA} and
  $k_{\text{\tiny 0}}=k_c(\mathbf{0})=0.0038$~\textup{\AA}$^{-1}$
  are the values of the linear coefficient and the momentum offset, respectively,
  in the absence of an external electric field.
It is noteworthy that 
  $\alpha(\mathbf{E})$ and $k_c(\mathbf{E})$ are both substantially enhanced
  with increasing $E$
  in the case of $\mathbf{E} \paralel \hat{\mathbf{x}}$,
  which exhibit a slight decrease for $\mathbf{E} \paralel \hat{\mathbf{y}}$.
The variation of $\alpha(\mathbf{E})$ with $E$ is seemingly \textit{superlinear}
  for $\mathbf{E} \paralel \hat{\mathbf{x}}$.
It is also notable that
  the (ZnO)$_{120}$:Bi nanowire
  has an \textit{anisotropic} response to the external electric fields
  as regards the degree of SO splitting of CB states.
Thus, in an experimental setup,
  it would be necessary first to determine the electric field directions
  for which $\alpha(\mathbf{E})$ and $k_c(\mathbf{E})$ show increasing and decreasing variations
  with increasing $E$.
This would enable directional control of the spin-split states
  in a practical application.

\begin{figure}
\centering
  \includegraphics[width=0.500\textwidth]{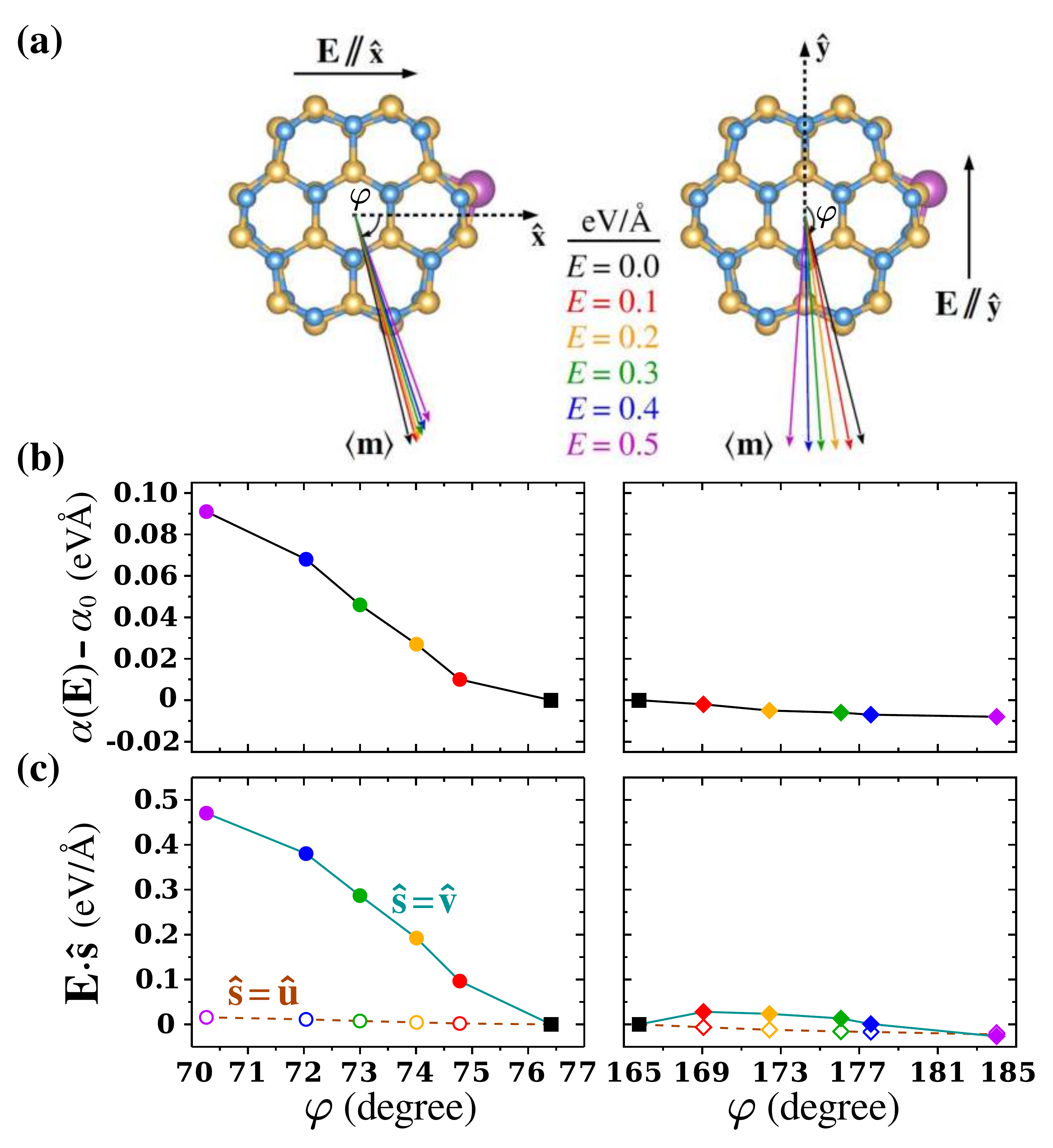}
  \caption{(a) The expectation value $\langle \mathbf{m} \rangle$ of the magnetization density with the CBM wave function
               for $\mathbf{E} \paralel \hat{\mathbf{x}}$ and $\mathbf{E} \paralel \hat{\mathbf{y}}$.
           The plots of (b) the electric-field-induced change in $\alpha$ and 
                        (c) the projections $\mathbf{E}\cdot\hat{\mathbf{s}}$
                        ($\hat{\mathbf{s}}=\hat{\mathbf{u}}$ and $\hat{\mathbf{v}}$)
                        with $\hat{\mathbf{s}} \perp \langle \mathbf{m} \rangle$
                        versus the angle $\varphi$ between $\mathbf{E}$ and $\langle \mathbf{m} \rangle$.
          }
  \label{f:cor}
\end{figure}

\begin{figure}
\centering
  \includegraphics[width=0.482\textwidth]{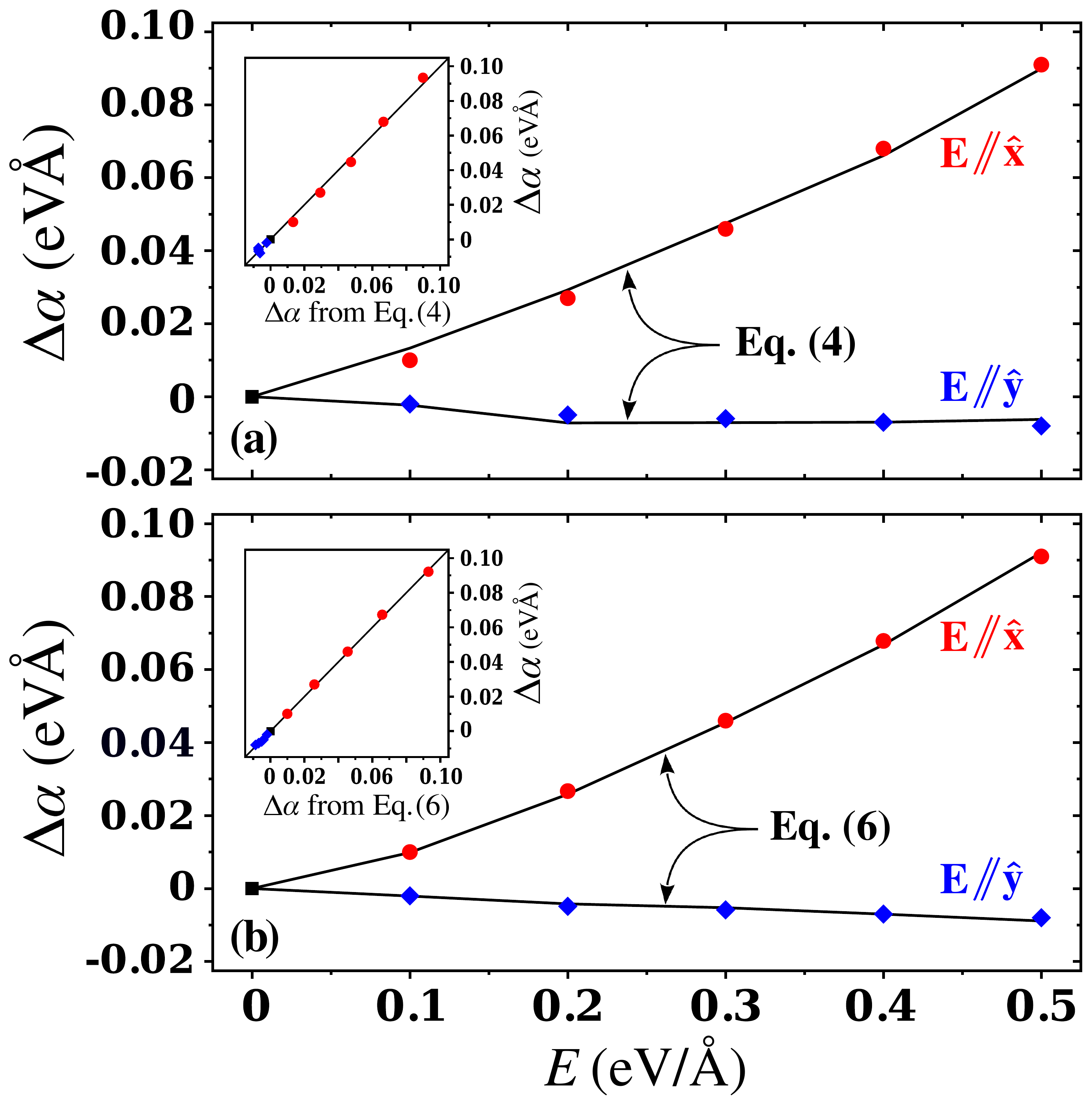}
  \caption{The plots of the electric-field-induced change $\Delta \alpha$ in $\alpha$ 
           versus the electric field strength $E$ for $\mathbf{E} \paralel \hat{\mathbf{x}}$ (the red  circles)
                                                  and $\mathbf{E} \paralel \hat{\mathbf{y}}$ (the blue diamonds).
          The black curves represent the result of fitting according to  (a) Eq.~(\ref{e:dae})
                                                                     and (b) Eq.~(\ref{e:dai}).
          The insets show the plots of the original $\Delta \alpha$ values
          versus the $\Delta \alpha$ values obtained from (a) Eq.~(\ref{e:dae})
                                                      and (b) Eq.~(\ref{e:dai}).
          }
  \label{f:delalf}
\end{figure}

\begin{figure}
\centering
  \includegraphics[width=0.500\textwidth]{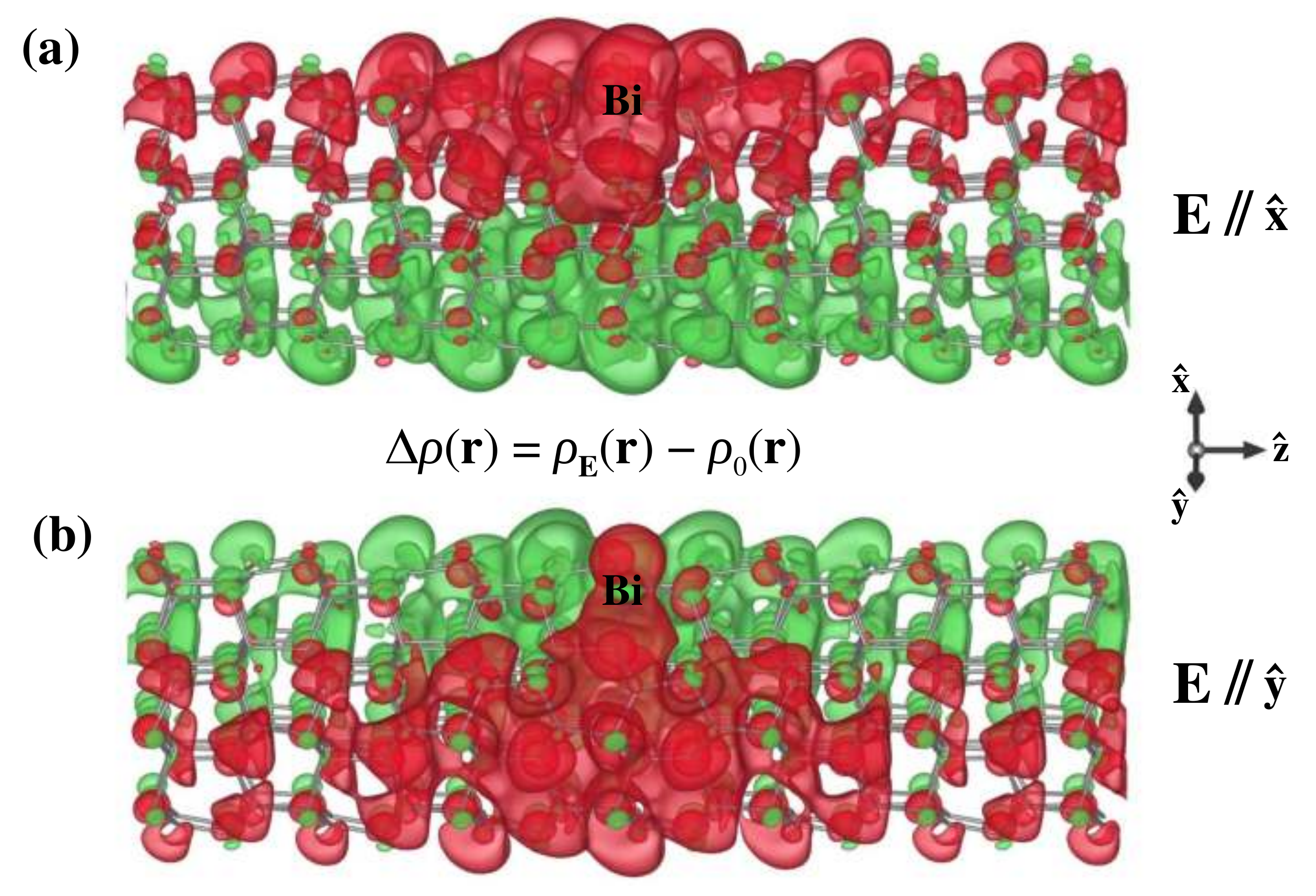}
  \caption{The isosurfaces of $\Delta \rho(\mathbf{r})$
             with isovalues of $\pm 5 \times 10^{-5}$~\textup{\AA}$^{-3}$
           for (a) $\mathbf{E}=0.5\hat{\mathbf{x}}$ and (b) $\mathbf{E}=0.5\hat{\mathbf{y}}$ eV/\textup{\AA}.
           The CBM state is \text{polarized} as imposed by the direction of the applied electric field
           since the red and green isosurfaces represent electric-field-induced increase and decrease in the CBM state charge density, respectively.
           }
  \label{f:delrho}
\end{figure}

From a fundamental physics point of view,
  the linear coefficient $\alpha$ introduced in Eq.~(\ref{e:disp})
  is related to the expectation value of the SO interaction operator
  $H_{\text{\tiny SO}}$ with $\psiE(\mathbf{r})$,
  which could be approximated \cite{aras17} as
  $\langle H_{\text{\tiny SO}} \rangle = -\langle \mathbf{m} \rangle \cdot \langle \mathbf B_{\text{\tiny SO}} \rangle$
  where
  $\langle \mathbf{m} \rangle$ and
  $\langle \mathbf B_{\text{\tiny SO}} \rangle$
  denote the expectation values of
  the magnetization density $\mathbf{m}(\mathbf{r})$ and
  the operator $\mathbf B_{\text{\tiny SO}} = -[(\vec{\nabla} V -e\mathbf{E}) \times \mathbf p]/2emc^2$, respectively.
A linear dependence of $\alpha(\mathbf{E})$ on $\mathbf{E}$ would therefore be expected.
Hence the nonlinear variation $\alpha(\mathbf{E})$ with the applied electric field
  in Fig.~\ref{f:var}(c) deserves further investigation.
To this aim, the variation of $\alpha(\mathbf{E})$ with $\mathbf{E}$
  will be analyzed in terms of 
  $\mathbf{E}$-induced changes in $\langle \mathbf{m} \rangle$
                              and $\langle \mathbf B_{\text{\tiny SO}} \rangle$.
Note that not only $\langle \mathbf B_{\text{\tiny SO}} \rangle$ but also $\langle \mathbf{m} \rangle$
  varies with $\mathbf{E}$
  in our noncollinear DFT calculations
  where $\mathbf{m}(\mathbf{r})$ is determined self-consistently,
  as shown in Fig.~\ref{f:cor}(a)
  where $\varphi$ denotes the angle between $\mathbf{E}$ and $\langle \mathbf{m} \rangle$.
The computed values for the magnitude of $\langle \mathbf{m} \rangle$
  are
  given in Table~S2 (see Ref.~\onlinecite{supmat}).
It should also be remarked that
  the projections of the vector $\mathbf{E}$ 
  \textit{perpendicular (parallel)} to the vector $\langle \mathbf{m} \rangle$
  make \textit{nonzero (zero)} contributions to $\langle H_{\text{\tiny SO}} \rangle$, and therefore to the splitting energy.
It is thus convenient to use an intrinsic coordinate system
  defined by the orthogonal unit vectors $\hat{\mathbf{u}}$, $\hat{\mathbf{v}}$ and $\hat{\mathbf{w}}$
  satisfying
  $\hat{\mathbf{w}} \paralel \langle \mathbf{m} \rangle$ and
  $\hat{\mathbf{u}}, \hat{\mathbf{v}} \perp \langle \mathbf{m} \rangle$,
  which could be taken as
  $\hat{\mathbf{u}}=\cos \theta \cos \phi \hat{\mathbf{x}}+\cos \theta \sin \phi \hat{\mathbf{y}}-\sin \theta \hat{\mathbf{z}}$,
  $\hat{\mathbf{v}}=           -\sin \phi \hat{\mathbf{x}}+            \cos \phi \hat{\mathbf{y}}$, and
  $\hat{\mathbf{w}}=\sin \theta \cos \phi \hat{\mathbf{x}}+\sin \theta \sin \phi \hat{\mathbf{y}}+\cos \theta \hat{\mathbf{z}}$,
  where $\theta$ and $\phi$ denote
  the angles between $\langle \mathbf{m} \rangle$ and $\hat{\mathbf{z}}$ and
                     $\langle \mathbf{m} \rangle$ and $\hat{\mathbf{x}}$, respectively.
For the $\langle \mathbf{m} \rangle$ vectors in Fig.~\ref{f:cor}(a),
  the angle pairs ($\theta$, $\phi$)
  take the 
  values given in Table~S2 (see Ref.~\onlinecite{supmat}).
It is important to notice that
  the electric-field-induced change in $\langle \mathbf B_{\text{\tiny SO}} \rangle$
  arising \textit{only} from the projections $\mathbf{E}\cdot\hat{\mathbf{s}}$ 
  with $\hat{\mathbf{s}} \ne \hat{\mathbf{w}}$
  yields a \textit{nonzero} contribution to $\langle H_{\text{\tiny SO}} \rangle$.
It is thus instructive
  to explore the relationship between 
  $\mathbf{E}$-induced variation of $\alpha$, i.e., $\Delta \alpha=\alpha(\mathbf{E})-\alpha_{\text{\tiny 0}}$,
  and the projections $\mathbf{E}\cdot\hat{\mathbf{s}}$ with $\hat{\mathbf{s}}=\hat{\mathbf{u}}$ and $\hat{\mathbf{v}}$
  as well as the $\mathbf{E}$-induced change in $\langle \mathbf{m} \rangle$,
  which will be denoted as  $\Delta \langle \mathbf{m} \rangle$.
A comparative inspection of Figs.~\ref{f:cor}(b) and \ref{f:cor}(c) reveals that
  the variation of $\Delta \alpha$ with respect to $\varphi$
  is of the same trend as that of $\mathbf{E}\cdot\hat{\mathbf{v}}$ ($\mathbf{E}\cdot\hat{\mathbf{u}}$)
  for $\mathbf{E} \paralel \hat{\mathbf{x}}$ ($\mathbf{E} \paralel \hat{\mathbf{y}}$).
In view of this and the foregoing discussion,
  a fitting according to
  \begin{equation}
    \Delta \alpha = c_u \mathbf{E}\cdot\hat{\mathbf{u}} + c_v \mathbf{E}\cdot\hat{\mathbf{v}} + c_m \Delta \langle \mathbf{m} \rangle \cdot \hat{\mathbf{w}}_{\text{\tiny 0}},
    \label{e:dae}
  \end{equation}
  was performed,
  where $\hat{\mathbf{w}}_{\text{\tiny 0}}$ denotes the unit vector in the direction of $\langle \mathbf{m} \rangle$
  in the \textit{absence} of the external electric field.
We found that the result of this fitting is \textit{not} entirely satisfactory,
  which was inferred from Fig.~\ref{f:delalf}(a).
We attribute the latter to the fact that
  the \textit{non-uniform} response of the nanowire
  to the applied electric field is \textit{not} taken into account in Eq.(\ref{e:dae}),
  which is demonstrated by
  the graphs of the $\mathbf{E}$-induced change $\Delta \rho(\mathbf{r})=\rhoE(\mathbf{r})-\rho_{\text{\tiny 0}}(\mathbf{r})$
  in the CBM state charge density
  in Fig.~\ref{f:delrho}.
The projections $\mathbf{E}\cdot\hat{\mathbf{s}}$ in Eq.(\ref{e:dae})
  are thus replaced by the integrals
  \begin{equation}
    I_s = \int -\vec{\nabla} \left [ \Delta V(\mathbf{r}) \right ] \cdot \hat{\mathbf{s}} \ \rhoE(\mathbf{r}) \ d^3r
    \label{e:iv}
  \end{equation}
  with $s=u$ or $v$,
  resulting in
  \begin{equation}
    \Delta \alpha = c_u I_u + c_v I_v + c_m \Delta \langle \mathbf{m} \rangle \cdot \hat{\mathbf{w}}_{\text{\tiny 0}}.
    \label{e:dai}
  \end{equation}
In Eq.(\ref{e:iv}),
  $\Delta V(\mathbf{r})$ denotes the electric-field-induced change in the \textit{self-consistent} potential $V(\mathbf{r})$.
Note that $I_s = \mathbf{E}\cdot\hat{\mathbf{s}}$
  for a \textit{uniform} electric field
  within a \textit{non-self-consistent} description.
The fitting according to Eq.(\ref{e:dai}) yields
  $c_u = 1.7464$~\textup{\AA}$^{2}$, $c_v = 0.3525$~\textup{\AA}$^{2}$, and $c_m = -0.3212$~\textup{eV\AA}/$\mu_{\text{\tiny B}}$,
  the result of which is quite satisfactory as seen in Fig.~\ref{f:delalf}(b).
The $\Delta \alpha$ values obtained from the right-hand-side (RHS) of Eq.(\ref{e:dai})
  are given in the second column of Table~\ref{t:da},
  which should be compared to the original values in the first column of the same table.
The contributions from the three terms in the RHS of Eq.(\ref{e:dai})
  are given in the third, fourth, and fifth columns of Table~\ref{t:da}. 
It is noticeable that the $c_u I_u$  and $c_v I_v$ terms have the greatest contribution to $\Delta \alpha$
  for $\Delta \alpha < 0$ and $\Delta \alpha > 0$, respectively,
  although the $c_m \Delta \langle \mathbf{m} \rangle \cdot \hat{\mathbf{w}}_{\text{\tiny 0}}$ term has also
  a non-negligible contribution.
Since both $c_u$ and $c_v$ are \textit{positive},
  a decrease in $\alpha$ occurs owing to $I_u < 0$ when $\mathbf{E} \paralel \hat{\mathbf{y}}$.
On the other hand, 
  an increase in $\alpha$
  occurs when $I_v > 0$ and/or $I_u > 0$.
Hence the increasing and decreasing variation of $\alpha$ with $\mathbf{E}$
  is traced to the \textit{sign} of the $I_s$ ($s=u$ and $v$) integrals,
  which is practically the same as the sign of $\mathbf{E}\cdot\hat{\mathbf{s}}$.
Accordingly, 
  as long as the directions $\hat{\mathbf{u}}$ and $\hat{\mathbf{v}}$ could \textit{not} be determined \textit{a priori},
  the direction of the applied electric field
  must be chosen carefully to ensure that an electric-field-induced \textit{increase} in $\alpha$ is achieved.
The latter would facilitate the directional control of the spin-split CB states, as also mentioned above,
  which would likely involve trial-and-error in a practical application.
It is nonetheless remarkable that
  a \textit{single} ZnO:Bi nanowire such as studied here
  could function as a spintronic device,
  operation of which is controlled by applying lateral electric fields.

\begin{table}
\centering
\caption{First and second columns:
         the electric-field-induced change in the linear coefficient $\alpha$
         computed from $\Delta \alpha=\alpha(\mathbf{E})-\alpha_{\text{\tiny 0}}$ and Eq.~(\ref{e:dai}), respectively.
         Third, fourth and fifth columns: 
         the contributions to $\Delta \alpha$ from the right-hand-side (RHS) terms of Eq.~(\ref{e:dai}).
         All values are in eV\textup{\AA}.
        }
\begin{tabular}{d{3.5} d{3.5} d{2.5} d{2.5} d{2.5}}
\hline\hline
 \multicolumn{1}{c}{$\Delta \alpha$} &
 \multicolumn{1}{c}{RHS of Eq.(6)} &
 \multicolumn{1}{c}{$c_u I_u$} &
 \multicolumn{1}{c}{$c_v I_v$} &
 \multicolumn{1}{c}{$c_m \Delta \langle \mathbf{m} \rangle \cdot \hat{\mathbf{w}}_{\text{\tiny 0}}$} \\
 \hline
 -0.008   & -0.009   &-0.0145 & -0.0035 & 0.0091  \\
 -0.007   & -0.007   &-0.0114 &  0.0001 & 0.0043  \\
 -0.006   & -0.006   &-0.0097 &  0.0015 & 0.0030  \\
 -0.005   & -0.004   &-0.0074 &  0.0026 & 0.0006  \\
 -0.002   & -0.002   &-0.0040 &  0.0018 & 0.0001  \\
  0.010   &  0.009   & 0.0014 &  0.0083 & 0.0002  \\
  0.027   &  0.026   & 0.0028 &  0.0213 & 0.0017  \\
  0.046   &  0.046   & 0.0048 &  0.0367 & 0.0040  \\
  0.068   &  0.067   & 0.0070 &  0.0522 & 0.0067  \\
  0.091   &  0.092   & 0.0100 &  0.0718 & 0.0113  \\
\hline\hline
\end{tabular}
\label{t:da}
\end{table}

In summary,
  the results of our density-functional calculations show that
  doping-induced linear-in-$k$ spin splitting of the lowest conduction-band states in a Bi-doped ZnO nanowire
  could be tuned by applying lateral electric fields
  via control of the electric field strength and direction.
We find
  that the degree of this splitting
  could be made to have a superlinear increase
  with increasing electric field strength,
  which is mediated by controlling the electric field direction.
Our analysis reveals that this is facilitated 
  by the non-uniform and anisotropic response of the ZnO:Bi nanowire
  to the applied electric field.
These findings indicate that
  a single ZnO nanowire doped with a low concentration of Bi
  could function as a spintronic device,
  operation of which is controlled electrically.

The authors acknowledge financial support from the Scientific and Technological Research Council of Turkey (TUBITAK) through Grant No. 114F155.
The calculations reported were carried out
  at the High Performance and Grid Computing Center (TRUBA Resources) of TUBITAK ULAKBIM.

\appendix*

\section{Bi defects in the ZnO:Bi nanowire}

\begin{figure}
\centering
  \includegraphics[width=0.482\textwidth]{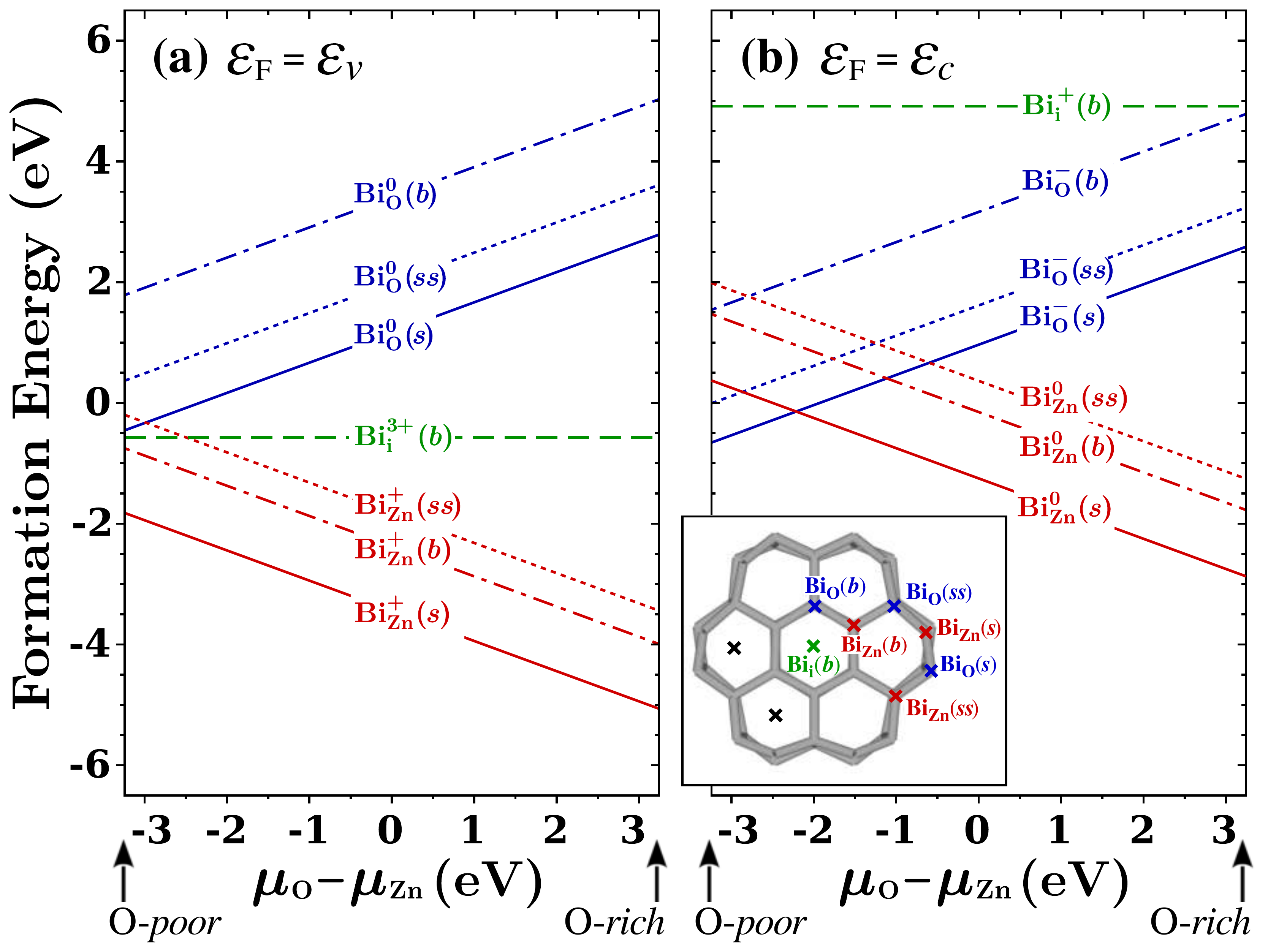}
  \caption{The formation energies of the extrinsic defects
           formed via the incorporation of Bi into the Zn, O or interstitial ($i$) sites
           in the bulk-like ($b$), surface ($s$) or subsurface ($ss$) regions of the ZnO nanowire,
           as shown in the inset,
           are plotted as a function of the difference $\mu_{\text{\tiny O}}-\mu_{\text{\tiny Zn}}$
           for (a) $\varepsilon_{\text{\tiny F}}=\varepsilon_v$ and
               (b) $\varepsilon_{\text{\tiny F}}=\varepsilon_c$
           where $\varepsilon_v$ and $\varepsilon_c$ denote the highest and lowest eigenvalue of the valence and conduction bands, respectively.
           The limiting values of $\mu_{\text{\tiny O}}-\mu_{\text{\tiny Zn}}$
           corresponding to O-\textit{poor} (i.e., Zn-\textit{rich}) and 
                            O-\textit{rich} (i.e., Zn-\textit{poor}) conditions are indicated by the vertical arrows.
           In the inset, the black cross marks represent the \textit{unlabeled} sites
           corresponding to the two discarded (unstable) configurations.
          }
  \label{f:forene}
\end{figure}

\begin{figure}
\centering
  \includegraphics[width=0.482\textwidth]{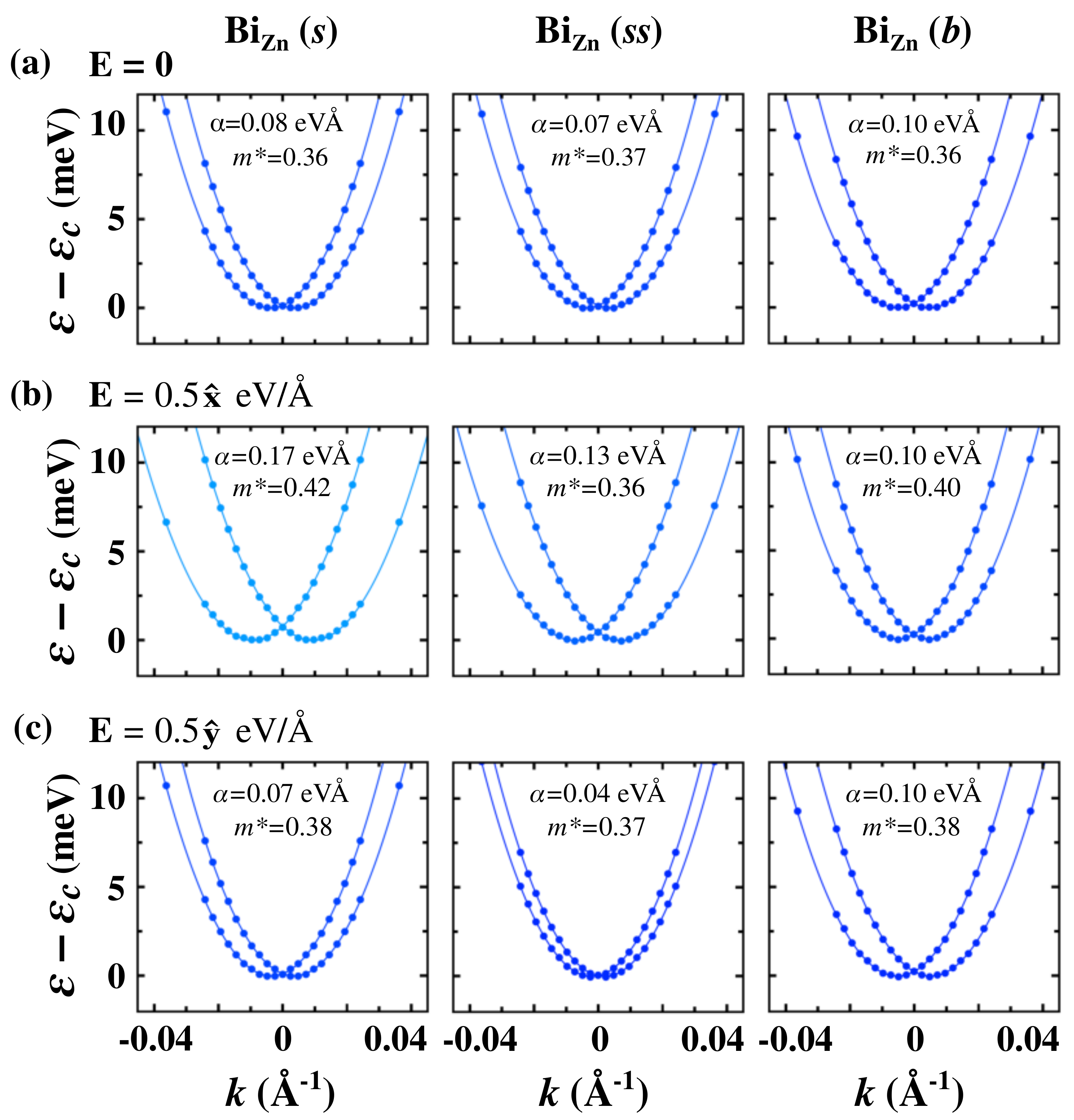}
  \caption{Spin-split conduction bands for ZnO:Bi nanowires containing $\text{Bi}_\text{Zn}(s)$,
                                                                       $\text{Bi}_\text{Zn}(ss)$, and $\text{Bi}_\text{Zn}(b)$
          for (a) $\mathbf{E}=\mathbf{0}$, (b) $\mathbf{E}=0.5\hat{\mathbf{x}}$~eV/\textup{\AA}, and (c) $\mathbf{E}=0.5\hat{\mathbf{y}}$~eV/\textup{\AA}.
          The values of the linear coefficient $\alpha$ (in eV\textup{\AA}) and the effective mass $m^{\ast}$ (in free-electron mass) are given in each panel.
          }
  \label{f:sssb}
\end{figure}

We have recently conducted \cite{kilic16} a theoretical characterization of a Bi-doped ZnO nanowire
  in a site-specific manner
  as regards the location and charge-state of the dopant,
  by calculating the defect formation energy $\Delta H_\text{f}$
  for a number of extrinsic defects
  formed via the incorporation of Bi into the Zn, O or interstitial ($i$) sites
  in the bulk-like ($b$), surface ($s$) or subsurface ($ss$) regions of the nanowire.
It is to be emphasized that $\Delta H_\text{f}$
  is an indicator for the {\it abundance} of the defect under given thermodynamic conditions
  since it is a significant portion of the {\it Gibbs energy of formation}
  that determines the {\it equilibrium defect concentration}.
The defects considered are shown in the inset of Fig.~\ref{f:forene},
  which are denoted as
  $\text{Bi}_\text{Zn}(b)$,
  $\text{Bi}_\text{O}(b)$,
  $\text{Bi}_i(b)$,
  $\text{Bi}_\text{Zn}(ss)$,
  $\text{Bi}_\text{O}(ss)$,
  $\text{Bi}_\text{Zn}(s)$,
  $\text{Bi}_\text{O}(s)$.
In structural optimizations,
  placing Bi initially at either of the \textit{unlabeled} sites shown in the inset of Fig.~\ref{f:forene}
  resulted in an \textit{unstable} configuration in which the nanowire is damaged.
These two unstable configurations are discarded.
In the present paper,
  the doping configuration displayed in Fig.~\ref{f:as}(a) contains the defect $\text{Bi}_\text{Zn}(s)$.
We studied the formation energies of the foregoing defects
  as a function of 
  the Fermi level $\varepsilon_{\text{\tiny F}}$ and 
  the atomic chemical potentials $\mu_{\text{\tiny Zn}}$, $\mu_{\text{\tiny O}}$ and $\mu_{\text{\tiny Bi}}$.
Our investigations \cite{kilic16,aras17} indicate that
  this doping configuration can be realized under reasonable thermodynamic conditions,
  which will be summarized here with the aid of
  the plots of  $\Delta H_\text{f}$ versus the difference $\mu_{\text{\tiny O}}-\mu_{\text{\tiny Zn}}$.
The latter are given
  in Figs.~\ref{f:forene}(a) and \ref{f:forene}(b)
  for \textit{two} limiting values of $\varepsilon_{\text{\tiny F}}$.
The value of $\mu_{\text{\tiny Bi}}$
  is set to the adsorption energy of a Bi atom on the nanowire surface, cf. Ref.~\onlinecite{aras17}.
It is seen in Figs.~\ref{f:forene}(a) and \ref{f:forene}(b) that
  the defect $\text{Bi}_\text{Zn}(s)$ in a charge state 0 or $+$ (depending on the location of the Fermi level)
  has not only \textit{lowest} but also \textit{negative} formation energies
  for a wide range of thermodynamic conditions.
The only exception to this is that
  ${\rm Bi}_{\rm O}(s)$ has a lower formation energy
  under O-\textit{poor} conditions for $\varepsilon_{\text{\tiny F}}=\varepsilon_c$
  in a narrow range of $\mu_{\text{\tiny O}}-\mu_{\text{\tiny Zn}}$.
Clearly, the formation of $\text{Bi}_\text{Zn}(s)$,
  rather than the rest of the alternatives with \textit{higher} formation energies,
  could be favored by adjusting thermodynamic conditions
  (i.e., by avoiding the values of $\mu_{\text{\tiny O}}$ and $\mu_{\text{\tiny Zn}}$
  corresponding to the latter range of $\mu_{\text{\tiny O}}-\mu_{\text{\tiny Zn}}$).
This means that the doping configuration displayed in Fig.~\ref{f:as}(a)
  could be realized under controlled thermodynamic conditions.
Besides, \textit{ab initio} molecular dynamics simulations performed
  at high temperature
  indicate that
  the same degree of stability could be assigned to the undoped ZnO and doped ZnO:Bi nanowires \cite{aras17}.
Finally, we think that the synthesis of surface-doped ZnO:Bi nanowire studied here
  would benefit from the low solubility of Bi in ZnO
  that derives the \textit{segregation} of Bi in ZnO varistors \cite{smith89, kobayashi98}.
A segregation tendency is also revealed for Bi in ZnO nanowires in Figs.~\ref{f:forene}(a) and \ref{f:forene}(b)
  where the surface ($s$) defects have lower formation energies compared to the respective bulk-like ($b$) and subsurface ($ss$) defects.

Since the formation energy of $\text{Bi}_\text{Zn}(s)$
  is significantly lower than that of $\text{Bi}_\text{Zn}(ss)$ and $\text{Bi}_\text{Zn}(b)$, cf. Figs.~\ref{f:forene}(a) and \ref{f:forene}(b),
  the equilibrium concentration of $\text{Bi}_\text{Zn}(s)$
  would be \textit{several orders of magnitude higher} than that of
  $\text{Bi}_\text{Zn}(ss)$ and $\text{Bi}_\text{Zn}(b)$ at room temperature.
It is nevertheless interesting to see 
  if spin splitting of the CB states occurs also for $\text{Bi}_\text{Zn}(ss)$ and $\text{Bi}_\text{Zn}(b)$.
Figures~\ref{f:sssb}(a)-(c) show the SO-split conduction bands for ZnO:Bi nanowires 
  containing $\text{Bi}_\text{Zn}(s)$,
             $\text{Bi}_\text{Zn}(ss)$, and $\text{Bi}_\text{Zn}(b)$
  for $\mathbf{E}=\mathbf{0}$, $0.5\hat{\mathbf{x}}$, and $0.5\hat{\mathbf{y}}$~eV/\textup{\AA}.
An expanded view of the band structures
  are provided in Fig.~S4 (see Ref.~\onlinecite{supmat})
  where the optimized atomic structures are included as insets.
Note that spin-orbit splitting of the CB states is present in each panel of Fig.~\ref{f:sssb}.
Moreover, the linear coefficient $\alpha$ takes quite similar values
  for $\text{Bi}_\text{Zn}(s)$, $\text{Bi}_\text{Zn}(ss)$ and $\text{Bi}_\text{Zn}(b)$.
On the other hand,
  $\alpha$ for $\text{Bi}_\text{Zn}(b)$ does not show much variation with $\mathbf{E}$.
The latter implies that electrical control of doping-induced spin splitting explored here
  could be achieved only in the case of surface (as opposed to bulk) doping of ZnO nanowires with Bi.

\clearpage

\section*{Supplemental Material}

\setcounter{page}{1}
\setcounter{figure}{0}
\setcounter{table}{0}
\renewcommand{\thepage}{S-\arabic{page}}
\renewcommand{\thefigure}{S\arabic{figure}}
\renewcommand{\thetable}{S\arabic{table}}

\begin{itemize}
\item Figure~S1 displays the equilibrium atomic configuration for the ZnO:Bi nanowire
    for the external electric fields of various strength applied in the \textit{x} or \textit{y} direction.
\item Table~S1 lists the electric-field-induced changes in the lengths of the bonds in Regions I and II depicted in Fig.~1(a).
Inspection of the values in this table reveals the following:
The O1$-$Bi (O2$-$Bi) bond exhibits the greatest shrinkage
  for $\mathbf{E} \paralel \hat{\mathbf{x}}$ ($\mathbf{E} \paralel \hat{\mathbf{y}}$),
  the degree of which is proportional to $E$.
In contrast, the O3$-$Bi bond that is roughly parallel to $\hat{\mathbf{z}}$
  changes relatively insignificantly and independent of $E$.
As shown in Fig.~1(a),
  the O1$-$Bi and O2$-$Bi bonds are located in Region I
  in the vicinity of the dopant.
The respective bonds in Region II (far from the dopant) are the O4$-$Zn7 and O5$-$Zn7 bonds
  that exhibit considerably smaller shrinkage.
Accordingly,
  the electric-field-induced structural changes are more pronounced in the vicinity of the dopant.
On the other hand,
  the O2$-$Zn1 and O3$-$Zn2 (O2$-$Zn1 and O3$-$Zn4) bonds exhibits the largest elongation
  for $\mathbf{E} \paralel \hat{\mathbf{x}}$ ($\mathbf{E} \paralel \hat{\mathbf{y}}$).
The O2$-$Zn1, O3$-$Zn2 and O3$-$Zn4 bonds are located in Region I.
The respective bonds in Region II are the O5$-$Zn8, O6$-$Zn9 and O6$-$Zn11 bonds
  that exhibit a similar degree of elongation.
In sum this analysis indicates that
  the electric-field-induced structural changes occur all around the nanowire,
  which are clearly more pronounced in the vicinity of the dopant.

\item Figure~S2 supplements Fig.~2.

\item Figure~S3 displays the electronic energy bands of the undoped ZnO nanowire 
           for $\mathbf{E}=\mathbf{0}$ and $\mathbf{E}=0.5\hat{\mathbf{x}}$~eV/\textup{\AA}.

\item Table~S2 lists the values for the magnitute and angles of the $\langle \mathbf{m} \rangle$ vectors depicted in Fig.~4(a).

\item Figure~S4 displays the electronic energy bands of ZnO:Bi nanowires
           containing substitutional defects $\text{Bi}_\text{Zn}(s)$,
                                             $\text{Bi}_\text{Zn}(ss)$, and
                                             $\text{Bi}_\text{Zn}(b)$
           for $\mathbf{E}=\mathbf{0}$, $0.5\hat{\mathbf{x}}$, and $0.5\hat{\mathbf{y}}$~eV/\textup{\AA}.
\end{itemize}

\clearpage

\begin{figure*}
\centering
  \includegraphics[width=1\textwidth]{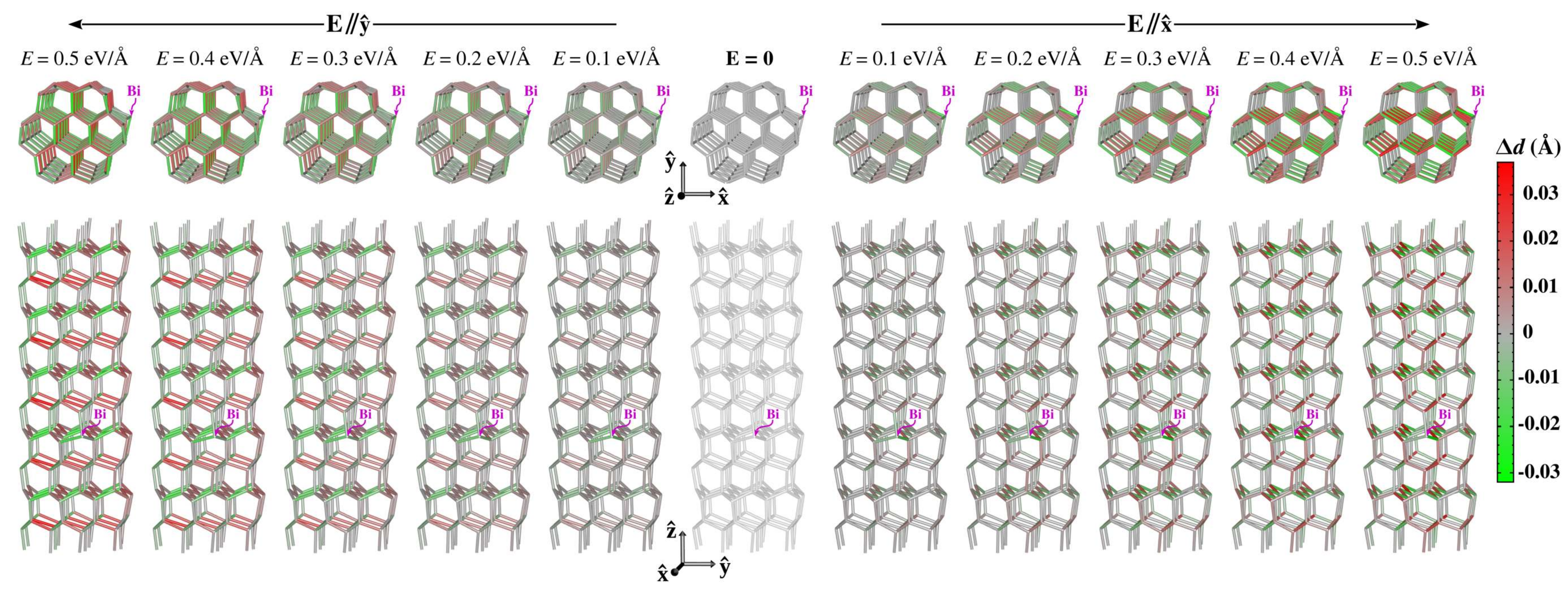}
  \caption{The equilibrium atomic configuration for the (ZnO)$_{120}$:Bi nanowire
           for $\mathbf{E}=$
               0.5$\hat{\mathbf{y}}$, 0.4$\hat{\mathbf{y}}$, 0.3$\hat{\mathbf{y}}$, 0.2$\hat{\mathbf{y}}$, 0.1$\hat{\mathbf{y}}$,
               $\mathbf{0}$,
               0.1$\hat{\mathbf{x}}$, 0.2$\hat{\mathbf{x}}$, 0.3$\hat{\mathbf{x}}$, 0.4$\hat{\mathbf{x}}$, and 0.5$\hat{\mathbf{x}}$~eV/\textup{\AA}.
           The sticks representing the Zn-O bonds are colored to reflect the electric-field-induced changes $\Delta d$ in the bond lengths.
          }
\end{figure*}

\begin{table*}
\centering
\caption{The electric-field-induced changes (in \textup{\AA}) in the lengths of the bonds in Regions I and II
         depicted in Fig.~1(a)
         for the external electric fields applied in the \textit{x} or \textit{y} direction.}
{\renewcommand{\arraystretch}{1.14}
\begin{tabular}{l d{-1.3} d{-1.3} d{-1.3} d{-1.3} d{-1.3} d{-1.3} d{-1.3} d{-1.3} d{-1.3} d{-1.3}}
  \hline\hline
            & \multicolumn{10}{c}{Electric field strength $E$ in eV/\textup{\AA}} \\ \cline{2-11}
            & \multicolumn{2}{c}{0.1} & \multicolumn{2}{c}{0.2} & \multicolumn{2}{c}{0.3} & \multicolumn{2}{c}{0.4} & \multicolumn{2}{c}{0.5} \\

  Bond      & \textit{x} &  \textit{y} &  \textit{x} &  \textit{y} & \textit{x} &  \textit{y} &  \textit{x} &  \textit{y} & \textit{x} &  \textit{y}   \\  \hline

  O1$-$Bi   &  -0.016 &  -0.009 &  -0.021 &  -0.007 &  -0.025 &  -0.005 &  -0.028 &  -0.004 &  -0.031 &  -0.003  \\
  O4$-$Zn7  &  -0.003 &  \hphantom{-}0.003 &  -0.007 &  \hphantom{-}0.004 &  -0.009 &  \hphantom{-}0.006 &  -0.012 &  \hphantom{-}0.008 &  -0.015 &  \hphantom{-}0.009  \\[4pt]

  O2$-$Bi   &  -0.005 &  -0.008 &  -0.005 &  -0.011 &  -0.005 &  -0.013 &  -0.004 &  -0.016 &  -0.004 &  -0.018  \\
  O5$-$Zn7  &   0.001 &  -0.002 &   0.003 &  -0.005 &   0.004 &  -0.007 &   0.005 &  -0.009 &   0.007 &  -0.011  \\[4pt]

  O3$-$Bi   &  -0.004 &  -0.005 &  -0.004 &  -0.004 &  -0.003 &  -0.004 &  -0.003 &  -0.004 &  -0.002 &  -0.004  \\
  O6$-$Zn7  &   0.001 &  \hphantom{-}0.000 &   0.003 &  \hphantom{-}0.001 &   0.004 &  \hphantom{-}0.001 &   0.005 &  \hphantom{-}0.001 &   0.007 &  \hphantom{-}0.001  \\[4pt]

  O2$-$Zn1  &   0.006 &  \hphantom{-}0.003 &   0.011 &  \hphantom{-}0.006 &   0.016 &  \hphantom{-}0.008 &   0.020 &  \hphantom{-}0.010 &   0.025 &  \hphantom{-}0.012  \\
  O5$-$Zn8  &   0.005 &  \hphantom{-}0.002 &   0.009 &  \hphantom{-}0.003 &   0.014 &  \hphantom{-}0.004 &   0.018 &  \hphantom{-}0.005 &   0.022 &  \hphantom{-}0.006  \\[4pt]

  O3$-$Zn2  &   0.007 &  \hphantom{-}0.001 &   0.011 &  -0.001 &   0.015 &  -0.001 &   0.019 &  -0.003 &   0.023 &  -0.004  \\
  O6$-$Zn9  &   0.005 &  -0.001 &   0.009 &  -0.002 &   0.014 &  -0.003 &   0.018 &  -0.004 &   0.022 &  -0.005  \\[4pt]

  O1$-$Zn3  &   0.003 &  -0.002 &   0.004 &  -0.006 &   0.005 &  -0.010 &   0.007 &  -0.013 &   0.008 &  -0.016  \\
  O4$-$Zn10 &   0.001 &  -0.004 &   0.001 &  -0.007 &   0.001 &  -0.010 &   0.001 &  -0.013 &   0.001 &  -0.016  \\[4pt]

  O3$-$Zn4  &   0.002 &  \hphantom{-}0.003 &   0.003 &  \hphantom{-}0.006 &   0.005 &  \hphantom{-}0.008 &   0.006 &  \hphantom{-}0.010 &   0.007 &  \hphantom{-}0.013  \\
  O6$-$Zn11 &   0.001 &  \hphantom{-}0.003 &   0.003 &  \hphantom{-}0.005 &   0.004 &  \hphantom{-}0.008 &   0.005 &  \hphantom{-}0.010 &   0.007 &  \hphantom{-}0.013  \\[4pt]

  O1$-$Zn5  &   0.006 &  \hphantom{-}0.004 &   0.007 &  \hphantom{-}0.004 &   0.008 &  \hphantom{-}0.004 &   0.008 &  \hphantom{-}0.003 &   0.009 &  \hphantom{-}0.003  \\
  O4$-$Zn12 &   0.001 &  \hphantom{-}0.000 &   0.000 &  \hphantom{-}0.000 &   0.000 &  \hphantom{-}0.000 &   0.000 &  \hphantom{-}0.000 &   0.000 &  \hphantom{-}0.001  \\[4pt]

  O2$-$Zn6  &   0.003 &  \hphantom{-}0.001 &   0.005 &  \hphantom{-}0.001 &   0.006 &  \hphantom{-}0.002 &   0.007 &  \hphantom{-}0.001 &   0.009 &  \hphantom{-}0.001  \\
  O5$-$Zn13 &   0.002 &  \hphantom{-}0.000 &   0.003 &  \hphantom{-}0.000 &   0.004 &  \hphantom{-}0.000 &   0.005 &  -0.001 &   0.007 &  -0.001  \\

  \hline\hline
\end{tabular}
}
\end{table*}

\begin{figure*}
\centering
  \includegraphics[width=1.000\textwidth]{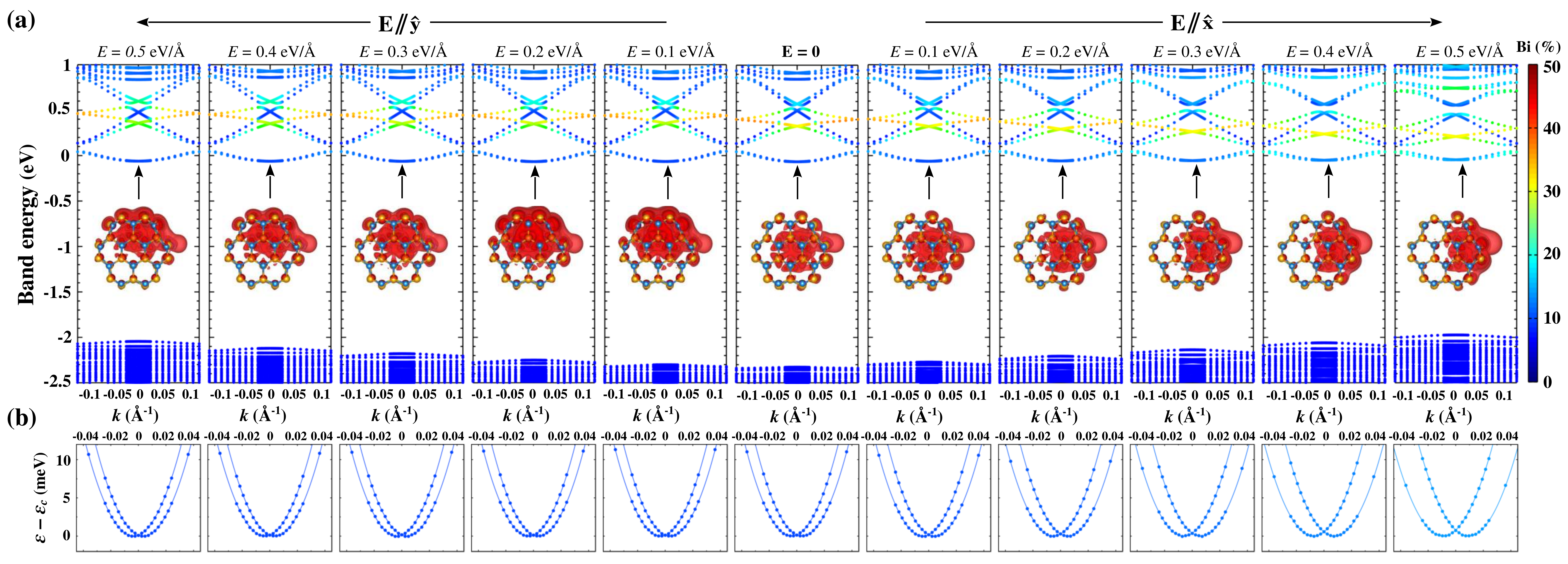}
  \caption{(a) The electronic energy bands of the (ZnO)$_{120}$:Bi nanowire under applied electric fields,
           which are colored to reflect the percent contribution from Bi to the electronic states.
           The Fermi level is set as the zero of energy.
           The vertical arrows point to the CBM.
           The red isosurfaces in the insets of the upper panels represent the CBM state charge densities for the isovalue of $1 \times 10^{-4}$~\textup{\AA}$^{-3}$.
           (b) Close-up views showing the spin-orbit splitting of the \textit{two} lowest conduction bands.
           }
\end{figure*}

\begin{figure*}
\centering
  \includegraphics[width=0.444\textwidth]{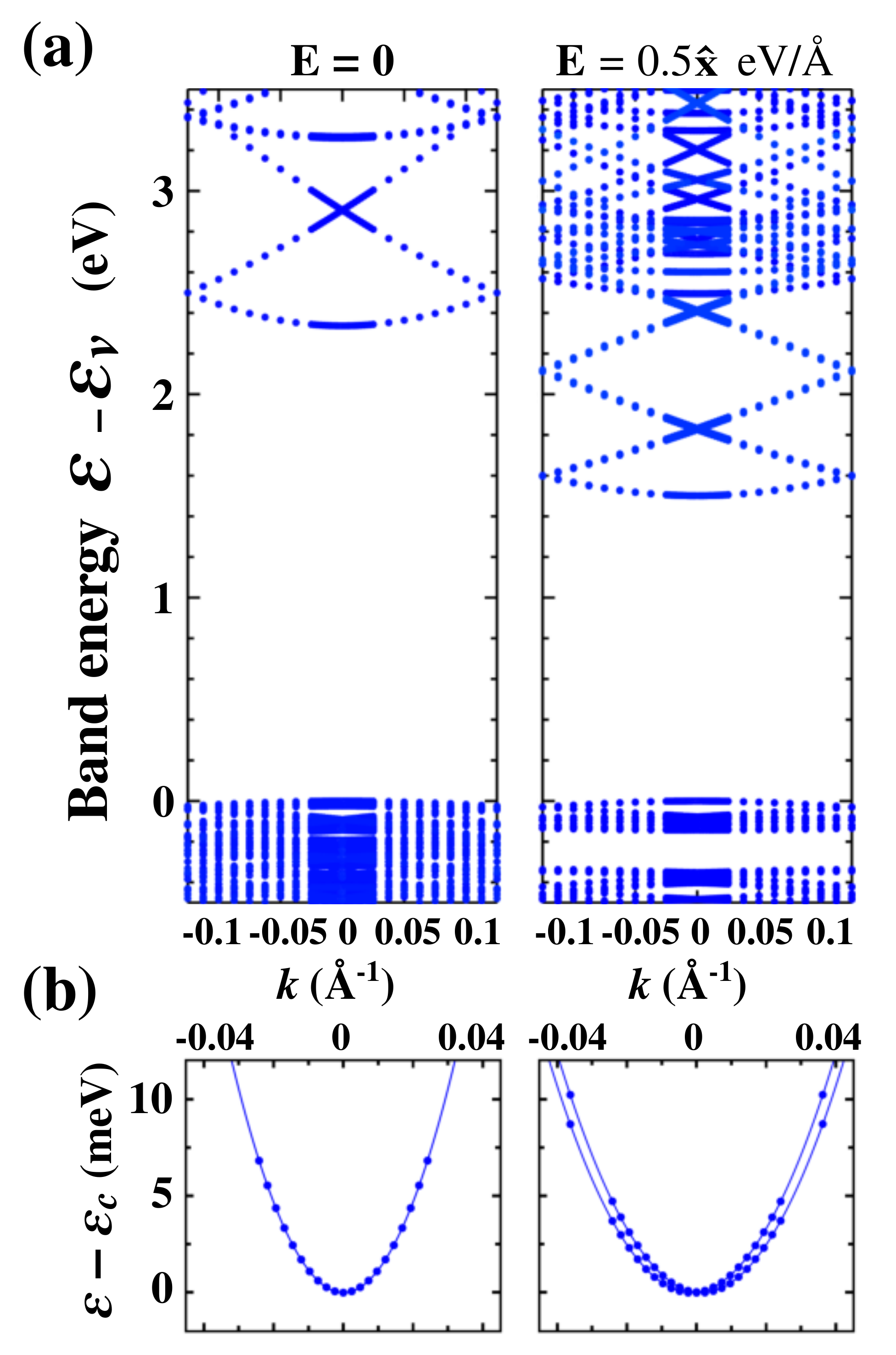}
  \caption{(a) The electronic energy bands of the \textit{undoped} (ZnO)$_{120}$ nanowire 
           for $\mathbf{E}=\mathbf{0}$ and $\mathbf{E}=0.5\hat{\mathbf{x}}$~eV/\textup{\AA}.
           The highest eigenvalue $\varepsilon_v$ of the valence band is set as the zero of energy.
           (b) Close-up views showing the \textit{two} lowest conduction bands.
           }
\end{figure*}

\onecolumngrid

\begin{table}
\centering
\caption{The values for the magnitute $\langle m \rangle$ (in Bohr magneton $\mu_{\text{\tiny B}}$) and $\theta$ and $\phi$ angles (in degrees)
         of the $\langle \mathbf{m} \rangle$ vectors depicted in Fig.~4(a)
         for given electric fields $\mathbf{E}$ (in eV/\textup{\AA}).}
{\renewcommand{\arraystretch}{1.14}
\begin{tabular}{ccccccc}
\hline\hline
$\mathbf{E}$          & ~~~ & $\langle m \rangle$ & ~ & $\theta$ & ~ & $\phi$  \\ \hline
$\mathbf{0}$          & ~~~ & 0.477               & ~ & 85.9     & ~ &  -76.4  \\
0.1$\hat{\mathbf{x}}$ & ~~~ & 0.476               & ~ & 85.7     & ~ &  -74.7  \\
0.2$\hat{\mathbf{x}}$ & ~~~ & 0.472               & ~ & 85.4     & ~ &  -73.9  \\
0.3$\hat{\mathbf{x}}$ & ~~~ & 0.465               & ~ & 85.1     & ~ &  -72.9  \\
0.4$\hat{\mathbf{x}}$ & ~~~ & 0.457               & ~ & 84.9     & ~ &  -72.0  \\
0.5$\hat{\mathbf{x}}$ & ~~~ & 0.444               & ~ & 84.7     & ~ &  -70.2  \\
0.1$\hat{\mathbf{y}}$ & ~~~ & 0.482               & ~ & 86.3     & ~ &  -79.7  \\
0.2$\hat{\mathbf{y}}$ & ~~~ & 0.478               & ~ & 86.5     & ~ &  -83.3  \\
0.3$\hat{\mathbf{y}}$ & ~~~ & 0.476               & ~ & 87.0     & ~ &  -87.5  \\
0.4$\hat{\mathbf{y}}$ & ~~~ & 0.476               & ~ & 87.6     & ~ &  -89.9  \\
0.5$\hat{\mathbf{y}}$ & ~~~ & 0.468               & ~ & 87.4     & ~ &  -93.1  \\
\hline\hline
\end{tabular}
}
\end{table}

\begin{figure*}
\centering
  \includegraphics[width=1.000\textwidth]{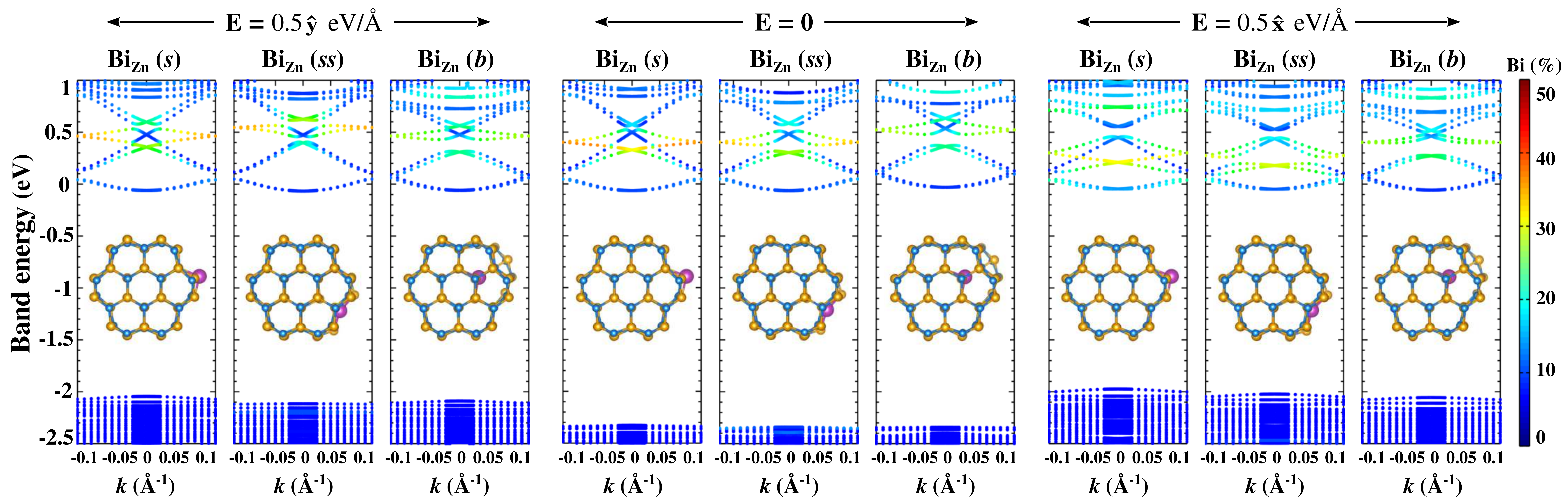}
  \caption{The electronic energy bands of (ZnO)$_{120}$:Bi nanowires
           containing substitutional defects $\text{Bi}_\text{Zn}(s)$,
                                             $\text{Bi}_\text{Zn}(ss)$, and
                                             $\text{Bi}_\text{Zn}(b)$
           for $\mathbf{E}=\mathbf{0}$, $0.5\hat{\mathbf{x}}$, and $0.5\hat{\mathbf{y}}$~eV/\textup{\AA},
           which are colored to reflect the percent contribution from Bi to the electronic states.
           The Fermi level is set as the zero of energy.
           Optimized atomic structures are shown as insets.
           }
\end{figure*}

\end{document}